\documentclass[usenatbib,useAMS]{mn2e}
\usepackage[english]{babel}

\usepackage[T1]{fontenc}
\usepackage{ae,aecompl}

\usepackage{fancyhdr}
\usepackage{amsfonts}
\usepackage{amsmath}
\usepackage{amssymb}
\usepackage{multicol}
\usepackage{layout}
\usepackage{graphicx}
\usepackage{multirow}

\usepackage{color}
\usepackage{multirow}
\usepackage[utf8]{inputenc}
\usepackage{times}
\usepackage{natbib}

\begin{document}

\title[Propagation of Electromagnetic Waves in MOG: Gravitational Lensing]{Propagation of Electromagnetic Waves in MOG: Gravitational Lensing}

\author[S. Rahvar and J. W. Moffat]
{S. Rahvar$^{1}$\thanks{rahvar@sharif.edu} J. W. Moffat$^{2,3}$\thanks{jmoffat@perimeterinstitute.ca}   \\
$^1$ Department of Physics, Sharif University of Technology, P.O.
Box 11155-9161, Tehran, Iran\\
$^2$ Perimeter Institute for Theoretical Physics, 31 Caroline St. N., Waterloo, ON, N2L 2Y5,Canada  \\
$^3$ Department of Physics and Astronomy, University of Waterloo,
Waterloo, Ontario N2L 3G1, Canada}

 \maketitle

\begin{abstract}

We investigate the solution of Maxwell's equations in curved spacetime within the framework of Modified Gravity (MOG). We show that besides the null-geodesic treatment of photons in MOG, using Maxwell's equations and covariant coupling
with the extra vector sector of gravitation in MOG, we can extract the equation for the propagation of light. We obtain Fermat's potential and calculate the deflection angle of light during lensing from a point-like star. Our results show that the deflection angle obtained from the solution of the wave equation in MOG for the large scale structures with larger impact parameter of light rays is proportional to that of General Relativity (GR). For solar mass stars the deflection angle agrees with the prediction of GR. However, for the compact structures like the supermassive black hole Sagittarius A* at the centre of the Milky Way, the prediction for the deflection angle is larger than GR, which can be tested in future observations.
\end{abstract}


\section{Introduction}

The observations of the dynamics of galaxies and clusters of galaxies reveal that the majority of
the mass of these structures, both in the framework of Newtonian gravity and General Relativity (GR), is claimed to be made of dark matter. The standard candidate for a dark matter particle is the Weakly Interactive Massive Particle (WIMP). The results of many experiments to detect WIMPs have failed to show evidence for the existence of dark matter particles~\cite{darkm}.

An alternative explanation of the dynamics of large-scale galaxy and galaxy cluster structures is replacing the dark matter with a modified gravity theory. The covariant theory of modified gravity (MOG) possesses a metric $g_{\mu\nu}$ and a vector field $\phi_\mu$, which are responsible for generating the gravitational field~\cite{moffat06}. In addition to the massless spin 2 graviton, the vector field $\phi_\mu$ is a massive spin 1 graviton field, which couples gravitationally to the current density of matter, and for a test particle results in a modified equation of motion. The extra term on the right-hand side of the equation of motion is a Lorentz-type gravitational force with a gravitational charge proportional to the mass of the test particle. The particles are in free fall satisfying the equivalence principle, but deviate from geodesic motion. In the weak gravitational field and slow motion of particles approximation, the field equations and the equation of of motion lead to a modified acceleration law.

It has been shown that MOG is consistent with the dynamics of galaxies and clusters of galaxies ~\citep{rahvar1,rahvar2,MoffatToth2,rahvar3,MoffatTothGalaxy,IsraelMoffat} without the need for dark matter and it is consistent with large scale cosmology data~\citep{MoffatCosmology,MoffatTothCosmology}. The theory can explain the gravitational lensing of galaxies and clusters of galaxies without the need for dark matter~\citep{brown,moglens,moglens2}. Moreover, MOG is consistent with the experimental result obtained from the neutron star merger event GW170617 that gravitational waves move with the speed of light~\cite{GreenMoffatToth}. In this work, we derive the gravitational lensing in MOG from the propagation of electromagnetic radiation within the background of a point mass star. The calculation is based on the coupling of the electromagnetic radiation field to the massive gravitational vector field $\phi_\mu$.

The paper is organized as follows: In Section (\ref{intmog}), we introduce the action and we discuss how the classical electromagnetic field can couple to the vector sector of MOG.  We investigate in Section (\ref{em_mog}), the propagation of the electromagnetic radiation in MOG and calculate the deflection angle due to single lensing and compare it to the null-geodesic equation in MOG. We end with conclusions in Section (\ref{conc}).

\section{Introduction to MOG}
\label{intmog}

The action in MOG is given by~\cite{moffat06}\footnote{In this work we use the signature of the metric $(+,+,+,-)$.} and we choose units with $c=1$:
\begin{equation}
\label{action1}
S=S_G+S_\phi+S_M,
\end{equation}
where
\label{Gravaction}
\begin{equation}
S_G=\frac{1}{16\pi}\int\frac{1}{G}\left(R+2\Lambda\right)\sqrt{-g}~d^4x,
\end{equation}
and
\begin{equation}
\label{B}
S_\phi=\frac{1}{4\pi}\int\Big(-\frac{1}{4}B^{\mu\nu}B_{\mu\nu}+\frac{1}{2}\mu^2\phi_\mu\phi^\mu + J_\mu\phi^\mu\Big)\sqrt{-g}~d^4x.
%\label{phi}
\end{equation}
Here, the Faraday tensor of the vector field $\phi_\mu$ is defined by
\begin{equation}
B_{\mu\nu}=\partial_\mu\phi_\nu-\partial_\nu\phi_\mu.
\end{equation}
\footnote{The dimensionless constant $\omega$ introduced in~\cite{moffat06} is set equal to unity}
The mass of the vector field $\phi_\mu$ is denoted by $\mu$, and $J^\mu$ is the matter current density coupled to $\phi_\mu$. For a perfect fluid we have
\begin{equation}
 J^\mu=\kappa\rho u^\mu,
\end{equation}
where $\kappa=\sqrt{\alpha G_N}$, $u^\mu=dx^\mu/ds$ and $G_N$ is Newton's gravitational constant. The gravitational charge that sources $\phi_\mu$ is
\begin{equation}
\label{vectorcharge}
Q_g=\int J^0d^3x=\kappa\rho u^0,
\end{equation}
where $\alpha$ is a dimensionless constant. We define the gravitational coupling in equation (\ref{Gravaction}) as $G=G_N(1+\alpha)$ where $\alpha$ is a constant parameter and can depend also to the mass of source of gravity. We will discuss later on details of this parameter.  

The equation of motion of a particle in MOG is given by
\begin{equation}
\label{Equatmotion}
m\biggl(\frac{du^\mu}{ds} + \Gamma^\mu{}_{\alpha\beta}u^\alpha u^\beta\biggr) = q_m B^\mu{}_\nu u^\nu,
%\label{Geo}
\end{equation}
where $\Gamma^\mu{}_{\alpha\beta}$ are the Christoffel symbols derived from $g_{\mu\nu}$, and $q_m=\kappa m=\sqrt{\alpha G_N}m$.
A derivation of the equation of motion from the MOG field equations and the conservation law $\nabla_\nu T^{\mu\nu}=0$, where $\nabla_\mu$ is the covariant derivative with respect to the metric $g_{\mu\nu}$, has been obtained by \citep{Roshan2}. For massive particles, we have
\begin{equation}
\frac{du^\mu}{ds} + \Gamma^\mu{}_{\alpha\beta}u^\mu u^\nu = \kappa B^\mu{}_\nu u^\nu.
\label{Geo}
\end{equation}
This demonstrates that particles in MOG fall freely in a homogeneous gravitational field independent of their composition, so MOG satisfies the weak equivalence principle, but the freely falling particles do not follow geodesic paths. For photons the photon mass $m_\gamma=0$ and $q_\gamma=\kappa m_\gamma=0$. It follows, that photons move along null geodesics:
\begin{equation}
\label{nullgeod}
k^\mu\nabla_\mu k^\nu =0,
\end{equation}
where $k^\mu$ is the photon 4-momentum null vector and $k^2=g_{\mu\nu}k^\mu k^\nu=0$. Gravitational radiation (gravitons) follows the same null geodesic as given by equation (\ref{nullgeod})~\cite{GreenMoffatToth}.

From the weak field and slow motion approximation, the MOG potential is given by
\begin{equation}
\label{ww}
\phi({\vec x}) = -G_N\int\frac{\rho({\vec x'})}{|{\vec x}-{\vec x'}|}\Bigl(1+\alpha - \alpha \exp(-\mu|{\vec x}-{\vec x'}|)\Bigr) d^3x'.
\end{equation}
The acceleration of a test mass particle is given by $a({\vec x}) = -{\vec \nabla}\phi({\vec x})$. The parameters $\alpha$ and $\mu$ are treated as constants in this potential. At small scales $|{\vec x}-{\vec x'}|\rightarrow 0$, the third term in the parenthesis approaches $\alpha$ and the potential becomes the Newtonian potential. At scales $|{\vec x}-{\vec x'}|\gg 1/\mu$, the MOG potential again becomes the Newtonian potential, but with an enhanced gravitational constant $G=G_N(1+\alpha)$.

For weak gravitational fields, the parameter $\alpha$ can be determined by the emipirical formula:
\begin{equation}
\label{alpha}
\alpha=\alpha_0\frac{M}{(\sqrt{M}+E)^2},
\end{equation}
where $\alpha_0$ and $E$ are constants. For the parameter $\mu$ we have
\begin{equation}
\label{mu}
\mu=\frac{D}{\sqrt{M}},
\end{equation}
where $D$ is a constant. The constants $D$ and $E$ are given by
\begin{equation}
D=6.25\times 10^3\,{M_{\odot}}^{1/2}\,{\rm kpc}^{-1},\quad E=2.54\times 10^4\,{M_{\odot}}^{1/2}.
\end{equation}
We choose $\alpha_0 = 10$, consistent with the best current estimate for the baryonic mass of the Milky Way, $M_b^{\rm MW}\sim 1.7\times 10^{11}~M_\odot$, which yields $\alpha^{\rm MW}=8.89$, the value used in our earlier study of Galaxy rotation curves and clusters~\citep{rahvar1,rahvar2,MoffatToth2,rahvar3}.

While the equation of motion in (\ref{Geo}) is used for massive particles, the null geodesic equation of motion of photons has been used to study the light deflection due to gravitational lensing. The result is an enhanced deflection of light, which compensates for the existence of dark matter in the galaxies and clusters of galaxies. In what follows, we recalculate the light deflection from the coupling of the electromagnetic field to the vector field $\phi_\mu$ and solve the wave equation.

The current density of matter is $J^\mu = \kappa\rho u^\mu$, which in terms of the energy-momentum tensor of a perfect fluid:
\begin{equation}
T^{\mu\nu}_{(M)}=(\rho+p)u^\mu u^\nu+pg_{\mu\nu},
\end{equation}
using $u^\mu u_\mu= -1$ can be written as
\begin{equation}
J^\mu = - \kappa u_\nu T^{\mu\nu}.
\end{equation}
We extend this definition to a generic matter field and use the subscript $M$ for it.  Now $T_{(M)}^{\mu\nu}$ can be associated with both a conventional fluid as well as fields such as the electromagnetic field, and the current density $J^\mu$ can be coupled to the vector field in the Lagrangian as $J^\mu\phi_\mu$.

\section{Propagation of the Electromagnetic Field in MOG}
\label{em_mog}
Let us consider the propagation of the electromagnetic field $F_{\mu\nu}$ in MOG in the vicinity of a point-like star. The matter Lagrangian density can be written as
\begin{equation}
L_{(M)} = \frac{1}{4\pi}\int\biggl( -\frac{1}{4} F^{\mu\nu}F_{\mu\nu}\sqrt{-g}\biggr)~d^4x + L_{S},
\end{equation}
where the first term is the electromagnetic field Lagrangian density and the second term is associated to a point-like star. We have
\begin{equation}
F_{\mu\nu} = \partial_\mu A_\nu - \partial_\nu A_\mu,
\end{equation}
where $A_\mu$ is the four-vector potential of the electromagnetic field. The variation of the Lagrangian density with respect to $A^\mu$ results in the electromagnetic field energy-momentum tensor:
\begin{equation}
T_{(EM)}^{\mu\nu} = \frac{1}{4\pi}(F^{\mu\alpha}F^{\nu}{}_\alpha -\frac{1}{4} g^{\mu\nu}F^{\alpha\beta}F_{\alpha\beta}).
\end{equation}

We introduce the coupling of the electromagnetic current vector $J_{\rm EM}^\mu=-\kappa k_\nu T^{\mu\nu}_{({\rm EM})}$ (in analogy to the perfect fluid) to the vector field, $ J^\mu\phi_\mu$, in the interaction term in equation (\ref{B}) as
\begin{equation}
\label{EMtensor}
J_{\rm EM}^\mu\phi_\mu = -\kappa k_\nu T^{\mu\nu}_{(EM)}\phi_\mu,
\end{equation}
where $k_\nu$ is the four-vector of electromagnetic radiation.  We also include the coupling of $\phi_\mu$ to the star:
\begin{equation}
J^\mu_{S}\phi_\mu=-\kappa u_\nu T^{\mu\nu}_{(S)}\phi_\mu,
\end{equation}
where $u^\mu$ is the four velocity of the star and $T^{\mu\nu}_{(S)}$ is the energy-momentum tensor of the star.

We rewrite the action of the electromagnetic field with the coupling to $\phi^\mu$ as follows:
\begin{eqnarray}
S_{EM} &=& \frac{1}{4\pi} \int \biggl( -\frac{1}{4} F^{\mu\nu}F_{\mu\nu} - \kappa\phi_\mu k_\nu F^{\nu\alpha} F^{\mu}{}_\alpha \nonumber \\
&+& \frac{1}{4}\kappa\phi_\alpha k^\alpha F^{\mu\nu}F_{\mu\nu}\biggr)\sqrt{-g} d^4x.
\end{eqnarray}
Varying this action with respect to $A^\mu$ results in the field equation:
\begin{eqnarray}
&& \biggl(\frac{1-\kappa k^\alpha \phi_\alpha}{\kappa}\biggr)\nabla_\mu F^{\mu\nu}{} =  -k_\alpha \phi^\mu\nabla_\mu F^{\alpha\nu} + k_\mu\phi^\nu\nabla_\alpha F^{\mu\alpha} \nonumber \\
&-& k^\alpha \phi_\mu \nabla_\alpha F^{\mu\nu} + k^\nu\phi_\mu\nabla_\alpha F^{\mu\alpha}
 - k_\alpha\nabla_\mu\phi^\mu F^{\alpha\nu} - k_\mu\nabla_\alpha\phi^\nu F^{\mu\alpha} \nonumber \\
 &-& k^\alpha{}\nabla_\alpha\phi_\mu F^{\mu\nu} + k^\nu\nabla_\alpha\phi_\mu F^{\mu\alpha} + k^\alpha \nabla_\mu \phi_\alpha F^{\mu\nu}.
\label{field}
\end{eqnarray}

We have assumed the weak gravitational field approximation for the propagation of the electromagnetic waves, and we expand the metric around the Minkowski metric $\eta_{\mu\nu}$:
\begin{equation}
g_{\mu\nu} = \eta_{\mu\nu} + h_{\mu\nu}.
\end{equation}
For the vector field we have
\begin{equation}
\phi^\mu = \phi^\mu_{(0)} + \phi^\mu_{(1)},
\end{equation}
and we assume in the flat Minkowski space that $\phi^\mu_{(0)} = 0$. In what follows, we drop for simplicity the index $(1)$ from the perturbation terms. For a point-like gravitational source with a time-independent static metric, the only non-zero component of $\phi_\mu$ is $\phi_0$.
We rewrite equation (\ref{field}) and ignore the second order perturbation terms:
\begin{eqnarray}
\label{nablaF}
\nabla_\mu F^{\mu\nu}{} &=& \kappa(-k_\alpha \phi^0\partial_0 F^{\alpha\nu} + k_\mu\phi^\nu\partial_\alpha F^{\mu\alpha} + k^\alpha \phi^0\partial_\alpha F^{0\nu} \nonumber \\
&-& k^\nu \phi^0 \partial_\alpha F^{0\alpha} - k_\mu \partial_\alpha\phi^\nu F^{\mu\alpha} +k^\alpha\partial_\alpha\phi^{0}F^{0\nu}  \nonumber \\
&-& k^\nu\partial_\alpha\phi^0 F^{0\alpha} -\partial_\mu \phi^{0} F^{\mu\nu}).
\end{eqnarray}

On the right-hand side of this equation, we keep only the ordinary derivatives as the connection terms result in second order perturbations, and the Latin letters represent the spatial components. Using the spatial component for $\nu$ on the left-hand side of (\ref{nablaF}) results in the wave equation:
\begin{eqnarray}
\label{nablaA}
&&\nabla^\mu\nabla_\mu A^{i} - \nabla^i\nabla_\mu A^{\mu} - R^{i}{}_{\mu}A^\mu =\kappa\phi^0(2\partial_0 F^{0i} + k^j\partial_j F^{0i} \nonumber \\
&-& k_j\partial_0F^{ij}+k^i\partial_jF^{0j})
+\kappa(k^i\partial_j\phi^0 F^{0i}-k^i\partial_j\phi^0F^{0j} \nonumber \\
&-&\partial_j\phi^0{}F^{ji}).
\label{we}
\end{eqnarray}
We impose the Lorenz gauge condition $\nabla_\mu A^\mu{} = 0$ and for empty space, we omit the third term on the left-hand side of equation (\ref{nablaA}).

The expression for $\nabla^\mu\nabla_\mu A^{i}$ in terms of partial derivatives with respect to $A^\mu$ and perturbations of the metric is derived in~\cite{rahvar2018} as follows:
\begin{eqnarray}
\label{11}
\nabla^\mu\nabla_\mu A^{i} & = &\partial^\mu\partial_\mu A^i - h^{\nu\alpha}\partial_\nu\partial_\alpha A^{i}{} + (\partial_\sigma h^{\alpha i}{} + \partial^\alpha h_\sigma{}^i{}\nonumber \\
&-& \partial^ih^\alpha{}_\sigma{})\partial_\alpha A^{\sigma}{} + \frac12(\partial_\nu\partial^\nu h_\sigma{}^{i}{}{} - \partial^\nu\partial_\nu h_\sigma{}^i \nonumber \\
&+& \partial_\nu\partial_\sigma h^{\nu i}{})A^\sigma - (\partial_\alpha h^{\alpha\sigma}{} + \frac12 \partial_\alpha h^{\alpha}{}^\sigma)\partial_\sigma A^i{},
\end{eqnarray}
where except the first term on the right-hand side, the rest of terms with the factor of perturbation of metric (i.e. $h^{\mu\nu}$) are the perturbation terms to the propagation of light in a flat space. However, we can argue that the rest of the perturbation terms are much smaller than this term. All the perturbation terms are multiplications of $A^\mu$ with the perturbation of the metric. Let us assume the length scale $L$ in the variation of the metric and the field $\phi_\mu$ and that $\lambda_e$ is the electromagnetic wavelength. Since $\lambda_e\ll L$, we ignore $(1/\lambda_e)(1/L)$ and $(1/L)^2$ terms compared to $(1/\lambda_e)^2$ terms. Then equation (\ref{11}) simplifies to
\begin{equation}
\label{result}
\nabla^\mu\nabla_\mu A^{i} = \partial^\mu\partial_\mu A^i - h^{\nu\alpha}\partial_\nu\partial_\alpha A^{i}{}.
\end{equation}
Following this argument for the perturbation terms of equation (\ref{nablaA}), we keep only the first term on the right-hand side of this equation. Then equation (\ref{nablaA}) simplifies to
\begin{equation}
\partial^\mu\partial_\mu A^i - h^{\nu\alpha}\partial_\nu\partial_\alpha A^{i}{} = \kappa\phi^0(2\partial_0 F^{0i} + k^j\partial_j F^{0i} - k_j\partial_0F^{ij}+k^i\partial_jF^{0j}),
\end{equation}
where in terms of the electromagnetic potential $A_\mu$, this equation can be written as follows:
\begin{eqnarray}
\partial^\mu\partial_\mu A^i - h^{\nu\alpha}\partial_\nu\partial_\alpha A^{i}{} &=& \kappa\phi^0(2\partial_0\partial^0 A^i - 2\partial_0\partial^iA^0 \nonumber \\
&-& k^j\partial_j\partial_i A^0 -k_j\partial_0\partial^i A^j).
\label{wave}
\end{eqnarray}
We can simplify the right-hand side of this equation using the Lorenz gauge. Also
assuming the first order solution of the wave equation, it follows that as $A^\mu(t) \propto {\cal A}_0^\mu e^{i\omega_0 t}$. Then, (\ref{wave}) simplifies to
\begin{equation}
\vec\nabla^2A^i + \Omega^2 A^i = -2\kappa \phi^0 \omega_0^2 \hat{k^i} A^0,
\label{wave2}
\end{equation}
where the effective frequency is $\Omega^2 = \omega_0^2(1+h^{00} + h^{ii} - 2\kappa \phi^0)$ and $\hat{k^i}$ is the unit vector along $A^i$. One of the features of (\ref{wave2}) is that $A^i$ is coupled to the $A^0$ term.  In conventional electromagnetism, this kind of coupling term is absent. Because $A^0$ in the perturbation term is an oscillating function, the final solution has an extra term containing $A^{0}$. The other feature of (\ref{wave2}) is that, for light rays far from the lens star, we can neglect $\phi^0$ as this term is exponentially damped, and the result is the same as the standard result of GR with the corresponding terms of the static, spherically symmetric MOG metric.

The static spherically symmetric metric around a star or a galaxy (or a black hole) in MOG is given by~\cite{moffat2015}:
\begin{eqnarray}
\label{metric}
ds^2 &=&-\left[1-2(1+\alpha)\phi_N + \phi_N^2\alpha(\alpha+1)\right]dt^2 \nonumber \\
&+& \left[1-2(1+\alpha)\phi_N + \phi_N^2\alpha(\alpha+1)\right]^{-1}dr^2 \nonumber \\
&+& r^2 d\theta^2+r^2 \sin^2\theta d\phi^2,
\label{mogm}
\end{eqnarray}
where $\phi_{N} = G_NM/r$ is the Newtonian potential.  For weak gravitational fields, we can determine the value of the constant parameter $\alpha$ from equation ({\ref{alpha}). For the strong gravitational fields of stellar mass MOG black holes, we must determine $\alpha$ from gravitational wave experiments. The $\phi^0$ term for a point-like mass has the solution $\phi^0(r) = -\kappa \exp(-\mu r)m/r$ ~\cite{moffat06,rahvar1}.  We note that for a point mass lens the spherically symmetric matter-free solution of MOG is a stationary solution of the MOG field equations with $T^{\mu\nu}=0$ (i.e matter-free), and satisfies the asymptotically flat Minkowski spacetime boundary condition. In this sense, it satisfies Birkhoff's theorem.  However, the weak field, slow motion approximation MOG acceleration formula of (\ref{ww}) does not satisfy the shell theorem or Gauss's law valid for a $1/r^2$ potential.In the  recent paper on the ultra diffuse galaxy NGC1052-DF2, we treat in detail how the shell theorem is modified and calculate the contribution from the Yukawa potential \cite{MoffatTothGalaxy}.

The equation (\ref{wave2}) has homogenous and non-homogenous parts to its solution. The non-homogenous part is
\begin{equation}
A_{non}^i(x,t) = \frac{\hat{k}^i \kappa\phi^0(x,t)}{\kappa\phi^0(x,t) - h^{00}} A^0(x,t),
\end{equation}
while the homogenous part of the solution of (\ref{wave2}) can be written as
\begin{equation}
{\vec\nabla}^2 A_h^i +  \Omega^2 A_h^i = 0.
\label{wave3}
\end{equation}
The effective frequency of the electromagnetic wave is given by
\begin{equation}
\label{omega0}
\Omega = \omega_0\left(1+2(1+\alpha)\phi_N - \alpha\exp(-\mu r)\phi_N\right).
\end{equation}
The solution of (\ref{wave3}) is given by the Kirchhoff integral:
\begin{equation}
A_h^i(r)=\frac{1}{4\pi} \int_S \biggl[A^i\frac{\partial}{\partial\hat{n}}\biggl(\frac{\exp(i\Omega s)}{s}\biggr) - \frac{\exp(i\Omega s)}{s}\frac{\partial A^i}{\partial\hat{n}}\biggr] dS,
\label{h}
\end{equation}
where the boundary of integration $S$ is taken close to the lens and $\Omega s$ represents the phase shift or the time delay of the different light rays received by the observer. The overall solution is the combination of the homogenous and non-homogenous solutions, $A(x,t) = A_h(x,t) + A_{non}(x,t)$, where at the location of the observer, far enough from the lens, since $\phi^0(x,t)$ decays exponentially, we can ignore the non-homogenous part of the solution. In what follows, we keep only the homogeneous part of the solution in equation (\ref{h}).

There is a standard approach in wave optics to derive the lensing equation by considering the interference of the light rays received by the observer at the screen $S$. We can calculate the geometric limit for
$\lambda_e \rightarrow 0$. Also, the trajectory of light in geometric optics is obtained from Fermat's potential~\cite{schneider,Nakamura,rahvar2018}. Let us define Fermat's potential as the overall phase of a light ray from the source to the observer.  Using equation (\ref{omega0}), it follows that
\begin{equation}
\Phi = \omega_0\int\Bigl(1+2(1+\alpha)\phi_N - \alpha \exp(-\mu r)\phi_N\Bigr)d\ell,
\end{equation}
where $\ell$ represents the path for the propagation of light.  We note that here we take lens as point like object and for a given lens, $\alpha$ is constant in this integral.  However, applying this integral for the lensing in the cosmological scales, $\alpha$ can depend on the scale and the mass of lens. The Fermat potential, considering the deflection of the light ray can be written as
\begin{equation}
\Phi = \frac{1}{2}D(\theta - \beta)^2 + 2(1+\alpha)\int\phi_N d\ell - \alpha\int\phi_N \exp(-\mu r)d\ell,
\label{fermat}
\end{equation}
where $D = D_{d}D_s/D_{ds}$, $\theta$ is the angular position of the image with respect to the lens-observer line of sight and $\beta$ is the angular position of the source in the absence of the lens.  Figure (\ref{fig1}) represents the configuration of the gravitational lensing. The integral of the second term on the right-hand side has an analytical solution. We denote by
$r= \sqrt{b^2 + \ell^2}$ the distance of the light ray from the source of gravity, where $b$ is the minimum impact parameter of the light ray.  We can replace it with $b = \theta D_{d}$, then equation (\ref{fermat}) simplifies to the following:
\begin{eqnarray}
\Phi &=& \frac{1}{2}D(\theta - \beta)^2 -4G_NM(1+\alpha)\ln\theta \nonumber \\
&-& G_NM \alpha \int \frac{\exp(-(b'^2 + \ell'^2)^{1/2})}{{(b'^2 + \ell'^2)}^{1/2}}d\ell',
\label{fermat3}
\end{eqnarray}
where variables with prime $'$ are normalized to the length scale of $1/\mu$. In what follows, our aim is to solve the integral in equation (\ref{fermat3}) and obtain the lens equation for different regimes.

{\it (i) Solar mass lens:}

For the small mass lens (on the order of solar mass $M_\odot$) from equation (\ref{alpha}), $\alpha \ll 1$,  and then varying the Fermat potential with respect to $\theta$ in equation (\ref{fermat3}) results in
\begin{equation}
%D(\theta - \beta) = \frac{4GM}{\theta}
\tilde{\theta} - \tilde{\beta} -\frac{1}{\tilde{\theta}} = 0,
\end{equation}
where all the angles are normalized to the Einstein angle with the definition of $\theta_E^2 = 4G_NM/D$. Here the deflection angle is given by $\alpha_{lens}  = 4G_NM/b$ and this result is in agreement with GR at the solar system scales. We note that for the solar mass lens from equation (\ref{alpha}) the value $\alpha\sim 10^{-9}$, guaranteeing that the MOG prediction for the bending angle for a photon grazing the limb of the Sun is $\Delta\phi=1.75''$ in agreement with GR and observations. All other predictions by MOG for the solar system, such as the perihelion advance of Mercury and the Cassini time delay observation will agree with experiment.
\begin{figure}
\begin{center}
\includegraphics[width=60mm]{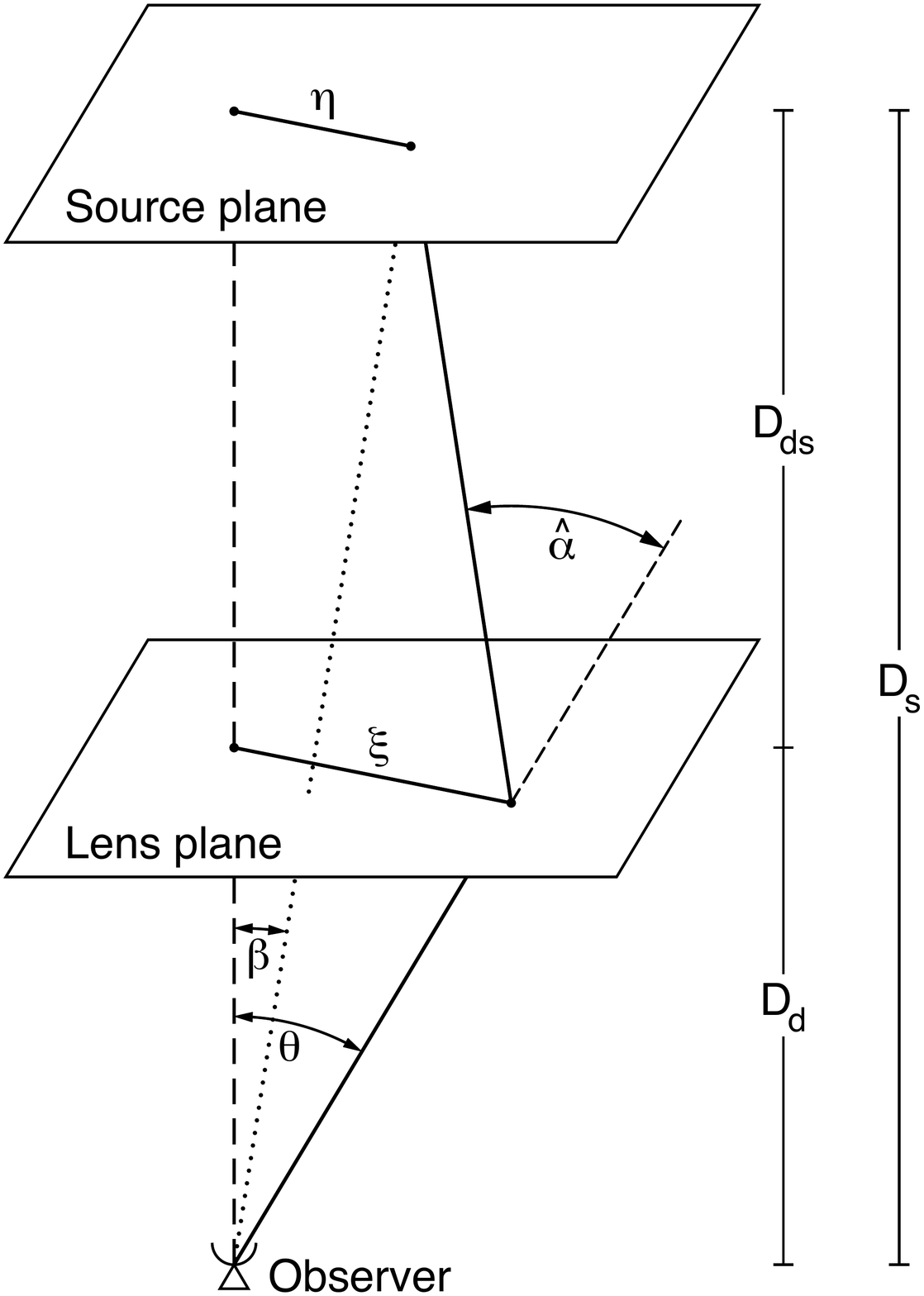}
\caption{ The configuration of gravitational lensing. The observer, lens and the source. The position of the source in the absence of the lens is observed with the angle $\beta$, and after lensing the angular position is $\theta$. The deflection angle is given by $\hat\alpha_{\rm lens}$. Here, $D_d$ is the distance of observer to the lens and $D_{ds}$ is the distance of the lens to the source. This figure is adapted from~\citep{Bart}.}
\label{fig1}
\end{center}
\end{figure}

{\it (ii) Lensing by a Galaxy: large impact parameter regime with $b\gg \mu^{-1}$}\\
We now investigate lensing with a larger mass comparable to a galaxy. For simplicity, we take light rays with a large impact parameter compared to $1/\mu$, then Fermat potential simplifies to
\begin{equation}
\Phi = \frac{1}{2}D(\theta - \beta)^2 -4G_NM(1+\alpha)\ln\theta - \sqrt{2} G_NM\alpha K_0\biggl(\frac{\mu D_d\theta}{2}\biggr),
\end{equation}
where $K_n$ is a Bessel function. Varying this equation with respect to $\theta$ results in the lens equation:
\begin{equation}
\tilde{\theta} - \tilde{\beta} -\frac{1+\alpha}{\tilde{\theta}} +\frac{\alpha\chi}{4\sqrt{2}}K_1(\chi\tilde{\theta}) = 0.
\label{r2}
\end{equation}
where $\chi = \mu D_d\theta_E/2$ and for the limit of $\chi\tilde{\theta}\gg 1$, which is the regime we are considering, the last term of equation (\ref{r2}) approaches zero and we obtain the lensing equation:
\begin{equation}
\tilde{\theta} - \tilde{\beta} -\frac{1+\alpha}{\tilde{\theta}} =0,
\end{equation}
where the deflection angle compared to GR is enhanced by the factor $(1+\alpha)$, which compensates for the existence of dark matter in galaxies and galaxy clusters. This result is consistent with the solution of the null geodesics equation.

{\it (iii) Lensing by a large compact object: small impact parameter regime with $b\ll \mu^{-1}$}\\
In this regime, the impact parameter is smaller than the characteristic size of the mass of the large compact object. 
%For a galaxy with the small impact parameter, the deflection angle is only influenced by the mass contained within the impact parameter distance scale, and hence the effective mass for the lensing would be smaller. 
For the small impact parameters equation (\ref{fermat3}) can be written as follows:
\begin{eqnarray}
\Phi &=& \frac{1}{2}D(\theta - \beta)^2 -4G_NM(1+\alpha)\ln\theta \nonumber \\
&-& G_NM \alpha \pi \Bigl(H_0(\mu D_d\theta) - Y_0(\mu D_d \theta)\Bigr),
\end{eqnarray}
where $H_n(x)$ is the solution of an inhomogeneous Bessel differential equation and $Y_n(x)$ is a Bessel function. 
For the small impact parameters the lens equation becomes:
\begin{equation}
\label{f2}
\tilde{\theta} - \tilde{\beta} -\frac{1+\alpha/2}{\tilde{\theta}} = \alpha\chi.
\end{equation}
For both stellar mass and large supermassive black holes the impact parameters of light are much larger than the Schwarzschild radii and we can use (\ref{alpha}) and $(\ref{mu})$ for weak gravitational fields. For a compact structure like an elliptical galaxy with the mass $M=10^9 M_\odot$, we obtain from (\ref{alpha}) and (\ref{mu}) the numerical values $\alpha =  3.08$, $\mu=0.19\,{\rm kpc}^{-1}$ and $\mu^{-1} = 5.06$\,{\rm kpc}. From $\chi=\frac{1}{2}R_E/\mu^{-1}$ and $D_d=0.5\,{\rm Gpc}$, we obtain $\chi = 0.02$ and we can ignore the right-hand side of (\ref{f2}). The lensing equation in this case is similar to the standard equation, where for our adopted value for the lens mass, the deflection angle is $\Delta\phi = 4G_NM/b(1+\alpha/2)\simeq 2.5\times4G_NM/b$. This predicts that for large supermassive compact objects the lensing results in a larger deflection angle.This deviation from GR can also be tested for the strong lensing systems where one of the images is close to the position of the lens (inside the Einstein ring) and the other one is outside the Einstein ring.

We can also examine in MOG the gravitational lensing by the central supermassive black hole Sagittarius A*~\cite{cbh}.  The mass is $M\simeq 4.1\times 10^6~M_\odot$. From (\ref{alpha}) and (\ref{mu}), the MOG parameters for this system are $\alpha = 0.055$, $\mu = 3.09\,{\rm kpc}^{-1}$ and $\mu^{-1}=0.32\,{\rm kpc}$, and $\chi\sim 4\times 10^{-5}$. These parameters simplify equation (\ref{f2}) to the standard gravitational lensing equation with minor corrections:
\begin{equation}
\tilde{\theta} - \tilde{\beta} -\frac{1.025}{\tilde{\theta}} = 0.
\end{equation}
From this equation, we get a $2.5\%$ deviation of the deflection angle compared to that obtained in GR. This result might be tested with the Event Horizon Telescope or future projects that study the light bending from the supermassive black hole at the center of the Milky Way.

\section{Conclusions}

We have postulated the form of the interaction of a fluid or the classical electromagnetic field $F_{\mu\nu}$ with the gravitational massive vector field $\phi_\mu$ in modified gravity (MOG) theory and solved Maxwell's field equations. The solution of the MOG electromagnetic wave equation for a point mass source resulted in a Kirchhoff integral with an effective electromagnetic wave frequency, where in the limit of zero electromagnetic wavelength, we recover the geometric optics limit and the light deflection due to gravitational bending. Our results at the solar system scale are in agreement with GR and for large scale structure the bending angle has an extra factor $1+\alpha$ for the bending of light: $\Delta\phi = 4\phi_N(1+\alpha)$ for the light rays with the impact parameter larger than $1/\mu$. This extra bending angle can compensate the effect of dark matter in the $\Lambda$CDM model.  We also studied lensing by massive compact objects with the impact parameter of the light smaller than $1/\mu$.  This case can happen when we have strong lensing around the bulge of a spiral galaxy or an elliptical galaxy with a small Einstein radius. In this case, the bending angle of light is $\Delta\Phi \simeq = 4\phi_N(1+\alpha/2)$ with masses in the range of $10^6 - 10^9~M_\odot$, where the extra factor produces an extra light bending angle.

Finally, we examined the gravitational lensing by a supermassive black hole at the center of a galaxy. Our results show that the bending angle of light in MOG is larger than GR. The observations of the central black hole Sagittarius A* in the Milky Way by the Event Horizon Telescope project or future observations that will identify the images of the lensing and may be used to test MOG.
\label{conc}

\section{Acknowledgments}

We would like to thank Viktor Toth for his helpful comments. Also we thank referee for his/her useful comments.This research was supported by Sharif University of Technology's Office of Vice President for Research under Grant No. G950214, and in part by Perimeter Institute for Theoretical Physics. Research at Perimeter Institute is supported by the Government of Canada through the Department of Innovation, Science and Economic Development Canada and by the Province of Ontario through the Ministry of Research, Innovation and Science.

\vspace{0.5cm}

%\section{References}
 %\begin{thebibliography}{}
%\bibitem[\protect\citeauthoryear{Afonso et al.}{2003}]{Af2003}
%\bibitem[{Afonso et al.}{2003}]{Af2003}

\bibliographystyle{mn2e}

% \bibliographystyle{jphysicsB}
% \bibliographystyle{harvard}

 % \bibliography{ref}

%\label{lastpage}

\end{document}